\documentclass[12pt]{iopart}

\usepackage{graphicx}

\usepackage{hyperref}

\hypersetup{
    colorlinks=true,
    linkcolor=blue,
    filecolor=magenta,
    urlcolor=violet,
    citecolor=blue,
   pdftitle={Sharelatex Example}
}

\begin{document}

\title[]{Impedance measurements in undoped and doped
regioregular poly(3-hexylthiophene)}

\author{Sougata Mandal$^*$ and Reghu Menon}

\address{Department of Physics, Indian Institute of Science, Bangalore 560012, India}
\ead{$^*$sougatam@iisc.ac.in}

\begin{abstract}
The semiconducting properties of regioregular poly(3-hexylthiophene) are characterized by impedance measurements, from 40 Hz to 100 MHz. X-ray diffraction shows the presence of both ordered and disordered regions. The analysis of impedance data by Nyquist plots show two semi-circular arcs, and its size is reduced by d.c. bias. Also, the carrier variation by light and chemical doping alters the shape and size of arcs. The fits to the data and equivalent circuits show considerable changes in the resistive, capacitive and constant-phase element parameters as the carrier density increases. The increase in carrier density reduces the relaxation time in ordered regions, and it does not alter much in disordered regions.
\end{abstract}


\maketitle

\section{Introduction}
The investigation of electrical properties of solution-processed semiconducting polymers has made considerable progress in past several years \cite{Le_2017}. In recent years semiconducting polymers are used in application like photovoltaic, field-effect transistors, photodiodes, printed electronics, etc  \cite{horowitz1990organic,clemens_fix_ficker_knobloch_ullmann_2004,ramuz2008high,choi2015development}. Nevertheless, the underlying charge transport mechanisms and its correlation to structure and morphology is yet to be fully understood \cite{kang2017charge}. It is well known that polymeric materials are intrinsically disordered, especially the presence of side-groups and its orientations alter the inter-chain interactions significantly \cite{doi:10.1021/ma981781f,doi:10.1021/ma951510u}. These factors modify both intra-chain and inter-chain delocalization of $\pi$ electrons, and this plays an important role in carrier mobility and device characteristics \cite{Khan_2017}. Disorder, traps, localization and mobility of carriers give rise to  wide range of charge transport mechanisms in semiconducting polymers. For example, Poole-Frenkel, injection-limited, space-charge limited, Fowler-Nordheim tunneling etc. have been reported \cite{doi:10.1063/1.3373393,Anjaneyulu_2011,HEEGER199423,doi:10.1063/1.1595707,doi:10.1063/1.3086882}. Apart from these, charge carrier density, trap density, molecular weight, etc. also observed to have important roles \cite{doi:10.1063/1.4775405,doi:10.1063/1.1891301}. All these factors control and limit the mobility in disordered semiconducting polymers. Poly(3-hexylthiophene) (P3HT) is a well known semiconducting polymer because of its easy solubility, stability and high mobility \cite{doi:10.1002/polb.24364,sirringhaus1999two}. Earlier studies in P3HT by photo-induced carrier generation indicate the presence of de-localize polarons \cite{osterbacka2000xm}. There are several studies related to d.c. charge transport of P3HT \cite{REP2003201,doi:10.1063/1.1805175}. As the usual d.c. transport studies by itself cannot unravel the transport and relaxation mechanisms, impedance measurements are quite useful in the characterization of semiconducting properties.

In this work, the resistive and capacitive contributions to charge transport in regioregular P3HT is being investigated by impedance measurements (40 Hz - 100 MHz). In undoped, photogenerated carriers, and chemically doped samples the impedance response are studied. This work is mainly focused on how the variation in carrier density affects the electrical transport in terms of relaxation mechanism in both order and disorder regions present in the sample and to quantify the resistive and capacitive elements by Nyquist plot.

\section{Experimental Details}

Regioregular P3HT (from Rieke Metals, Inc.) is used in this study \cite{doi:10.1021/ja00106a027}. The films are prepared by drop-casting from a solution of chlorobenzene, and thickness of free-standing films is around $20\mu m$. X-ray diffraction measurement is performed by Rigaku SmartLab high-resolution X-ray diffractometer (XRD) in grazing angle (GI angle = $0.5^\circ$) mode (GIXRD). The instrument is calibrated with a standard silicon sample and operated at 40 kV/30 mA with CuK$_\alpha$ ($\lambda=1.54A ^\circ$) radiation from $2\theta$ = $3^\circ$ to $80^\circ$ and diffraction pattern is recorded by scintillation counter detector. For impedance studies, electrical contacts are made by carbon paint on front and back side of sample. Impedance measurements are performed by precision impedance analyser (Agilent 4294A) from the frequency range 40 Hz to 100 MHz under d.c. bias 0V and 3V with a.c. signal amplitude of 50 mV. Impedance measurements are performed in dark and light conditions. Light condition measurements are performed using continuous illumination of blue light, that is above the band-gap of the sample. For studies in chemically doped samples, FeCl$_3$ in nitromethane solvent is used as dopant, and the doping levels are monitored by measuring the $in$-$situ$ variation in resistance of sample while doping. In this way, two doped samples are prepared: a very lightly doped sample (doping-1) in which the resistance decreased only by a factor of two (from 80 kohm to 40 kohm) and in the moderately doped sample (doping-2) the resistance decreased from 40 to 2 kohm, in the same sample. As it is known that in conducting polymers at high doping levels the resistance typically decreases by many orders of magnitude, the doping levels in the samples in this work is on the lower side.

\section{Results and Discussions}

\begin{figure}[htp]
\centering
\includegraphics[scale=0.4]{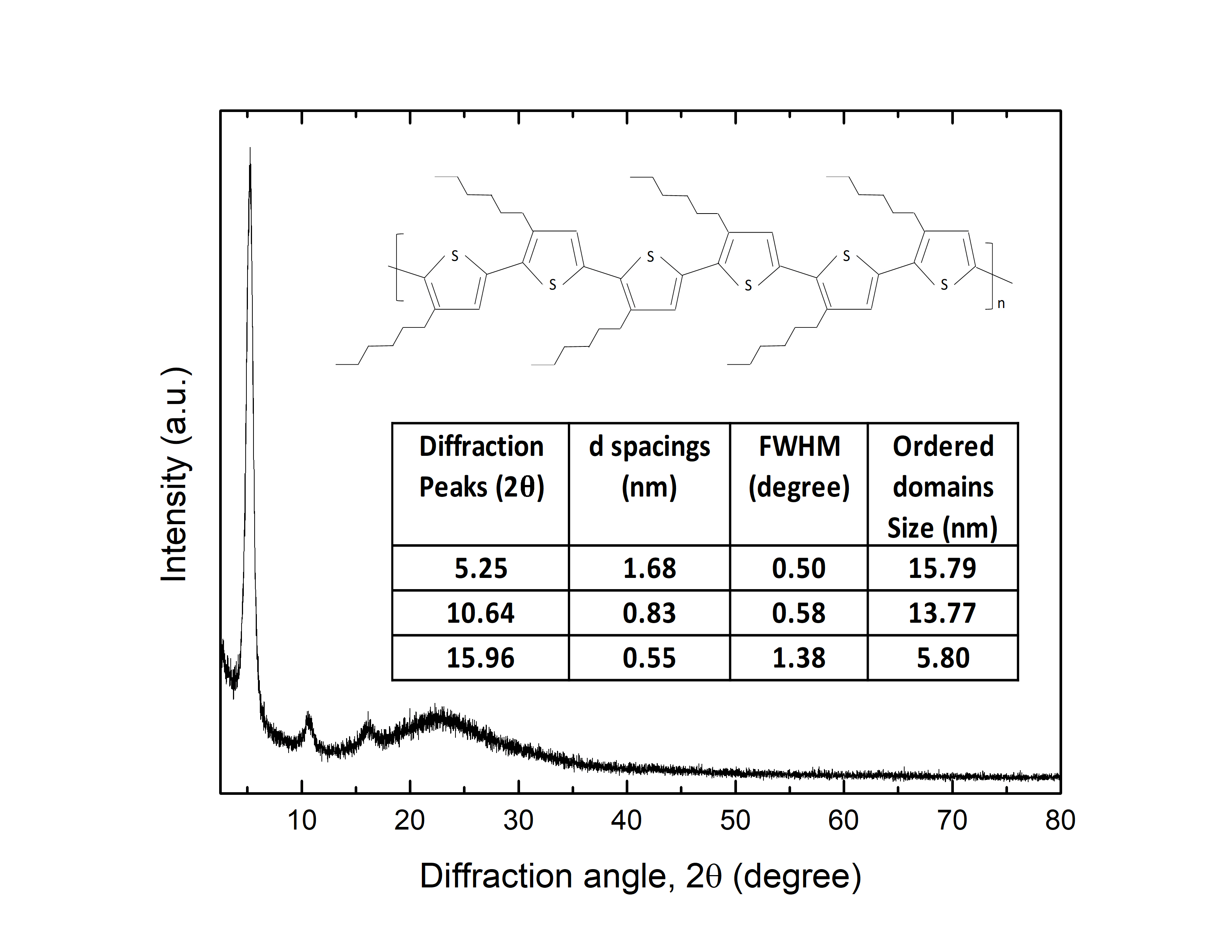}
\caption{XRD pattern of regioregular poly (3-hexylthiophene), inset shows the structure and the data.}
\label{fig.1}
\end{figure}

The grazing-incidence (0.5$^\circ$) X-ray diffraction of free-standing regioregular P3HT films is shown in figure \ref{fig.1}. The pattern shows one sharp peak at 2$\theta$ = 5.25$^\circ$ and two smaller peaks at 10.64$^\circ$ and 15.96$^\circ$ and a broad
hump around 2$\theta$ = 22.6$^\circ$. As reported before, these peaks are attributed to (100), (200) and (300) planes which signify the well-organized domains due to stacking of thiophene rings as expected in regioregular structure of polymer chains with ordered hexyl-groups \cite{2006NatMa...5..197K}. Peak intensity signifies the extent of well-organized domains \cite{2006NatMa...5..197K,doi:10.1002/adfm.200400521}. The broad hump around 2$\theta$ = 22.6$^\circ$ is mainly due to scattering from disordered hexyl groups and tilts in thiophene rings \cite{WINOKUR1989419}. The size of domains (t) and d-spacings are calculated using the equations [$t = 0.9\lambda/\beta \cos\theta; 2d\sin\theta = n\lambda$, where $\lambda$ is X-ray wavelength, $\beta$ is line broadening at half of the maximum intensity (FWHM), $\theta$ is Bragg angle, $n$ is the order of reflection] as shown in the inset of figure \ref{fig.1}. These values are rather similar to that in earlier reports \cite{doi:10.1002/adfm.200400521,WINOKUR1989419,cullity1978elements}. Hence, the X-ray data show that the quality of free-standing film of P3HT is quite good.

\begin{figure}[htp]
\centering
\includegraphics[scale=0.5]{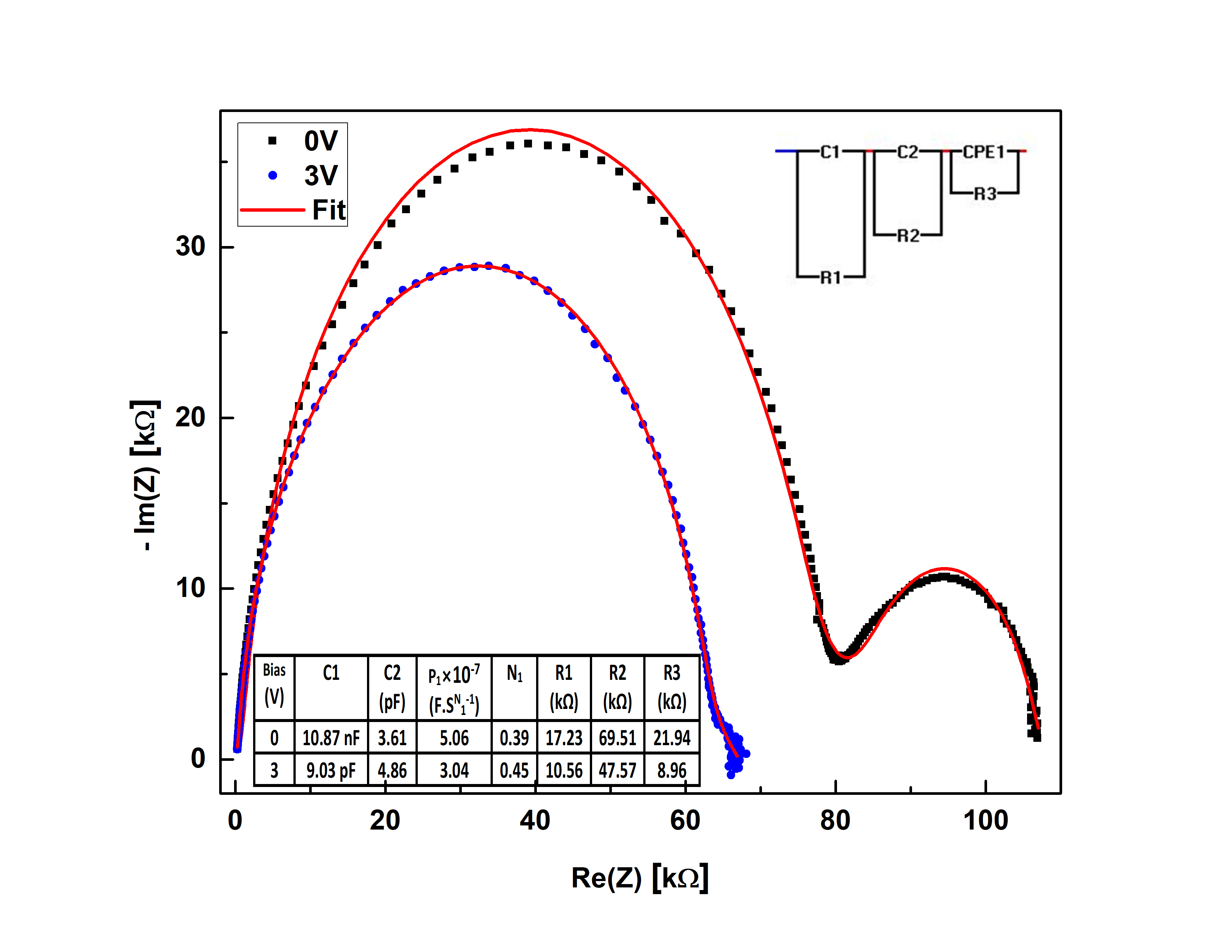}
\caption{Nyquist plots for P3HT at different voltages in dark and fitted with equivalent circuit as shown in inset and the table shows the fit values to equation (\ref{equ 1}).}\label{fig. 2}
\end{figure}

It is known that regioregular P3HT is a well-ordered semiconducting polymer with room temperature conductivity values of the order of 10$^{-6}$ to 10$^{-8}$ S/cm, depending on the absorption and quasi-doping by oxygen \cite{PhysRevB.80.195211}. The well-ordered domains are due to stacking of thiophene rings and $\pi-\pi$ interactions; and disordered parts mainly from tilt in rings and randomness of hexyl groups. Complex impedance measurements, as a function of frequency and bias voltage, are a useful method to probe resistive and capacitive contributions to probe the overall transport, unlike in case of d.c. transport. Previous impedance studies in conducting polymers like polypyrrole (PPy) have shown that distribution of relaxation times depends on doping; and relaxation frequency also depends on  disorder as the relaxation time of charge carriers varies with disorder or traps in the sample \cite{doi:10.1063/1.4775405,mostany1997impedance,Varade_2013}. Usually in impedance ($Z$) studies a combination of real [resistance ($R$)] and imaginary part [consisting of both capacitive and inductive contributions ($X$)], as expressed by $Z = R + jX$ is observed. The frequency dependence of resistive part is mainly due to polarizations within the sample. The capacitive part arises mainly due to accumulation of charges in traps and domain boundaries. Reactance of capacitive part and inductive part are 1/$j\omega$C and $j\omega$L, respectively. Typically the ac response of data is analyzed by fitting to Cole-Cole [or Cole-Davidson or Havriliak-Negami] relaxation equation \cite{Varade_2013,kao2004dielectric}. The equivalent circuit obtained from fit parameters gives the dissipation factor, loss factor and relaxation time constant \cite{kao2004dielectric}. In polymeric materials these factors dependents on morphology, traps, charge carrier density and mobility.

\begin{figure}[htp]
\centering
\includegraphics[scale=1]{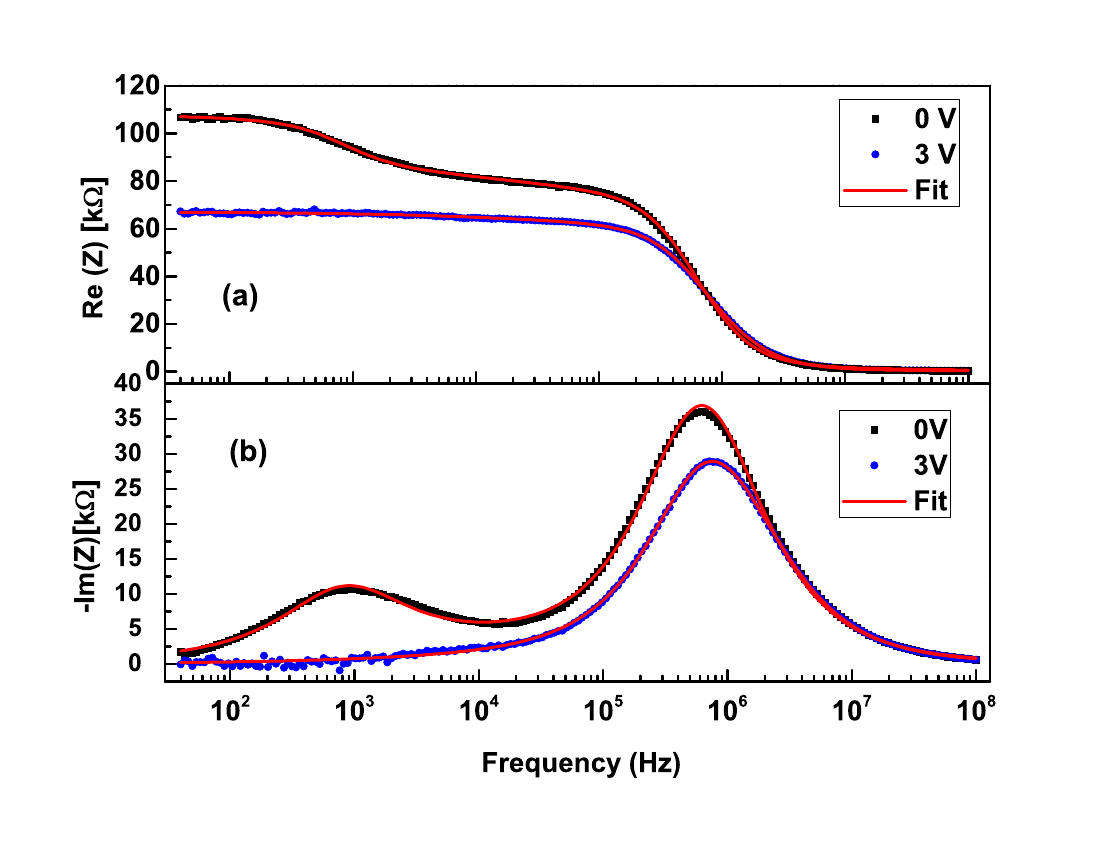}
\caption{(a) Real part of impedance vs frequency (b) Imaginary part of impedance vs frequency in dark. Data fits to equations (\ref{equ 2}) and (\ref{equ 3}).}\label{fig. 3}
\end{figure}

The complex impedance data for undoped P3HT in dark is shown in figure \ref{fig. 2}. The zero bias voltage data show two semi-circular arcs; the larger one is at higher frequencies and the smaller one is at lower frequencies. The former one is mainly due to the faster relaxation process occurring at the relatively more ordered regions; and the later one is arising from larger time constants involved in relaxations  frequency in  disordered  regions of the sample. After applying a bias voltage of 3 V, the peak value of higher frequencies semi-circular arc is reduced as well as the lower frequencies impedance response is barely observable due to the injection of charge carriers. The injected carriers diffuse rapidly, and it tends to reduce the large-scale variations in the local electrical properties of sample. This fact is more evident when the real and imaginary parts are plotted as a function of frequency as shown in figures \ref{fig. 3}. The complex impedance data fitted to equation (\ref{equ 1}) as given by
\begin{eqnarray}
Z&=&\frac{R1}{1+j\omega C1R1}+\frac{R2}{1+j\omega C2R2}+\frac{R3}{1+{(j\omega)}^{N_1}P_1R3}\label{equ 1}
\\
Re(Z)&=&\frac{R1}{1+\omega^2{C1}^2{R1}^2}+\frac{R2}{1+\omega^2{C2}^2{R2}^2} \nonumber \\
&+&\frac{R3\left(1+R3P_1\omega^{N_1}\cos{\frac{N_1\pi}{2}}\right)}{\left(1+R3P_1\omega^{N_1}\cos{\frac{N_1\pi}{2}\ }\right)^2+\left(R3P_1\omega^{N1}\sin{\frac{N_1\pi}{2}}\right)^2}\label{equ 2}
\\
-Im(Z)&=&\frac{{\omega C1R1}^2}{1+\omega^2{C1}^2{R1}^2}+\frac{{\omega C2R2}^2}{1+\omega^2{C2}^2{R2}^2}\nonumber \\ &+&\frac{{R3}^2P_1\omega^{N_1}\sin{\frac{N_1\pi}{2}}}{\left(1+R3P_1\omega^{N_1}\cos{\frac{N_1\pi}{2}\ }\right)^2+\left(R3P_1\omega^{N_1}\sin{\frac{N_1\pi}{2}}\right)^2}\label{equ 3}
\end{eqnarray}
where, Z is the total impedance; $\omega$ is the angular frequency; $R1$, $R2$, $R3$ are the resistive elements;$C1$, $C2$ are capacitive elements; $P$1 is a constant and exponent $N1$ describes the capacitive nature of constant phase element (CPE) parameter. $Re(Z)$ and $Im(Z)$ are real and imaginary part of impedance($Z$), respectively. The obtained values of resistive and capacitive elements along with the equivalent circuit, are shown in inset of figure (2). For fitting, we tried several circuits and choose the best fit to yield quite realistic values for resistance and capacitance. The fits are not arbitrary and trivial as from the obtained values of resistance and capacitance values by varying the carrier density. For e.g., as the number of carriers increases the resistance values should go down. From the fit elements of impedance data, there are two R-C element and one $R$-CPE. The impedance of CPE is given by $1/P(j\omega)^ N$. Each one of the $R-C$ circuit corresponds to one of  the semicircle in Nyquist plot \cite{cole1941dispersion}. The smaller semicircle at lower frequencies represents the overall contributions from $R2-C2$ and $R3$-CPE1. Since the value for $R1$ is much lower than that of $R2$ and $R3$, the former is associated with more ordered regions in sample, as displayed by the large semicircle at higher frequencies. The combined contributions from $R2-C2$ and $R3$-CPE1, as represented in the smaller semicircle at lower frequencies is associated with more disordered regions in the sample; hence larger values for both $R2$ and $R3$, also the randomness in trap states can give rise to a lower value for $C2$. When 3V bias voltage is applied, the injected carriers reduce the values for resistive elements and C1; as already trapped charges are present in disordered regions, $C2$ does not vary much under bias voltage. The increased carrier density at 3V bias voltage nearly smears out the smaller semicircle at  lower  frequencies  due  to less significant contribution to transport from disorder regions, as expected in more ordered regioregular P3HT samples. The frequency dependence of both real and imaginary parts are fitted using equations (\ref{equ 2}) and (\ref{equ 3}) as shown in figure \ref{fig. 3}. At zero bias voltage  the slope changes twice at lower frequencies and the values drop significantly around 10$^{6}$ Hz; however, it becomes nearly flat at lower frequencies at 3 V bias voltage and drop in values occur at similar frequency range. These features corroborate well with frequency dependence of imaginary part as in figure 3(b). Hence the presence of two relaxation process, at zero bias voltage, indicates the variations in carrier transport and the associated time-scales involved in ordered and disordered regions in sample. However, the transport at lower frequencies is modified by 3 V bias voltage, as the injected carriers tend to average out the relaxation processes at longer time-scale; and at higher frequencies there is no significant variations.

\begin{figure}[htp]
\centering
\includegraphics[scale=0.5]{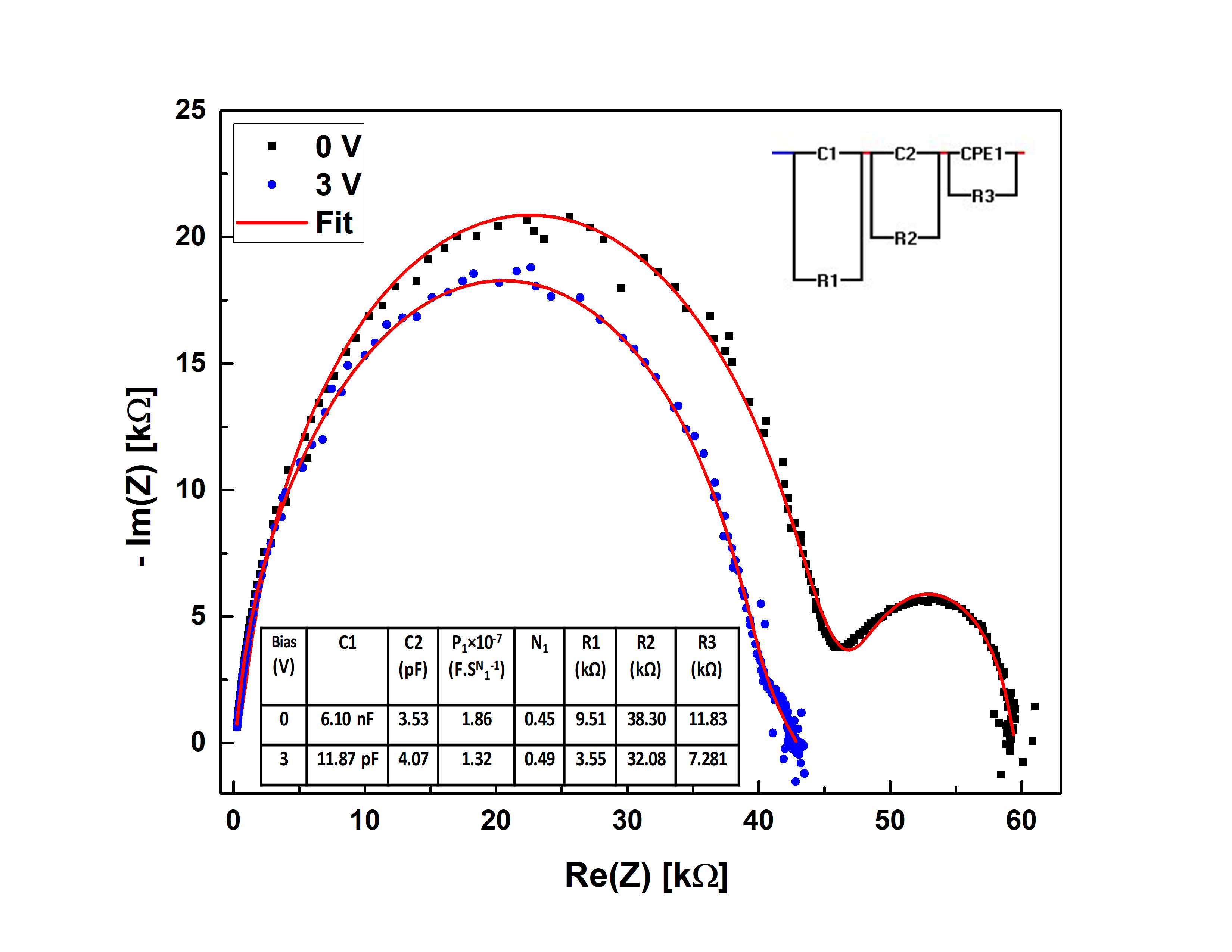}
\caption{Nyquist plots for P3HT at different voltages in light and fitted with equivalent circuit as shown in inset and the table shows the fit values to equation (\ref{equ 1}).}\label{fig. 4}
\end{figure}

\begin{figure}[htp]
\centering
\includegraphics[scale=1]{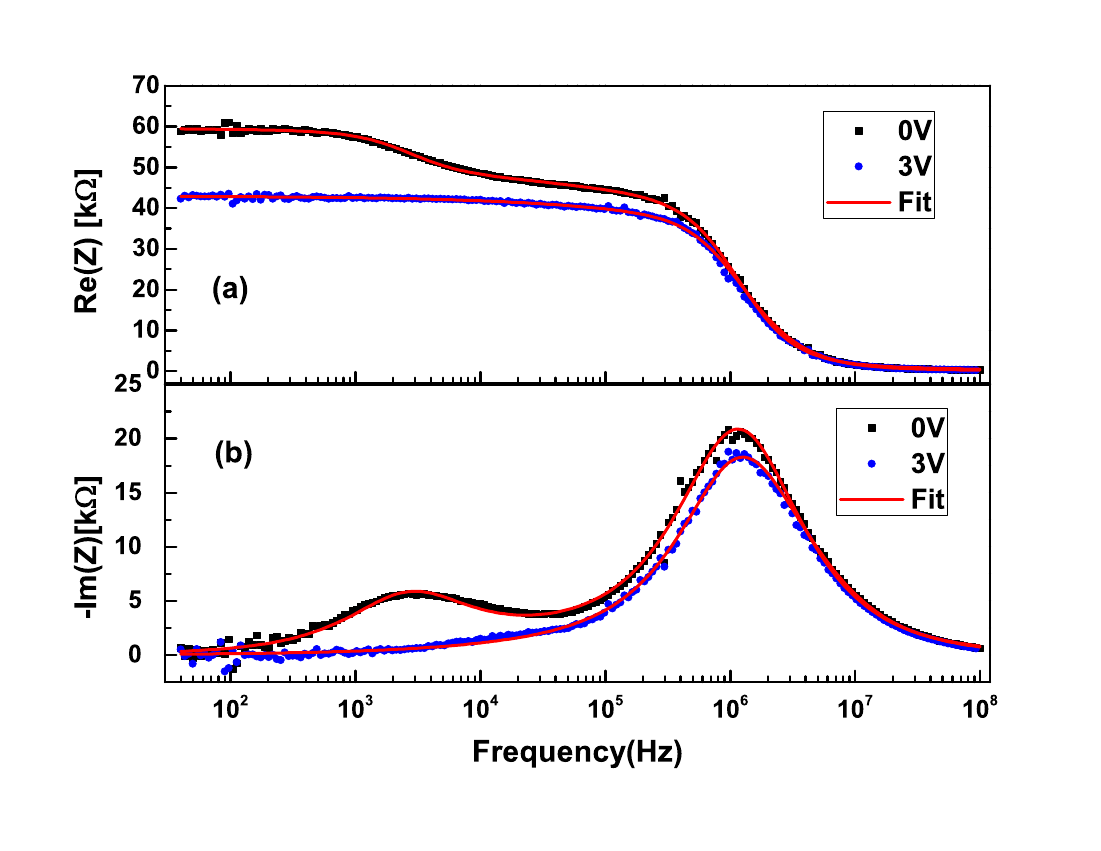}
\caption{(a) Real part of impedance vs frequency (b) Imaginary part of impedance vs frequency in light. Data fits to equations (\ref{equ 2}) and (\ref{equ 3}).}\label{fig. 5}
\end{figure}

Since P3HT is a semiconductor polymer with bandgap around 2.2 eV, it absorbes light at wavelength below 562 nm, and photo-generation of carriers occur by shining with blue light. These photocarriers are expected to modify the electrical response of material. The complex impedance data for photo-doped P3HT is shown in figure \ref{fig. 4}. The basic features are rather similar as in dark condition. The photo-generated carriers at different dc biases reduces both capacitance and resistance values. Obtained values for the fit from equation (\ref{equ 4}), as in the inset of figure \ref{fig. 4}, show variations from that of the dark measurement.  The injected  carriers at 3 V bias  voltage  further  reduces these values, hence the combined effect of both photo-doped carriers and injected carriers can be observed. The frequency dependence of real and imaginary parts of the impedance data, upon photo-doping, is shown in figure \ref{fig. 5}, and it fit to equations (\ref{equ 2}) and (\ref{equ 3}). The trends are nearly similar to that observed under the dark condition, with variations in the fit-parameter values.

The time-constant $\tau$ values for $R-C$ ($\tau = RC$) and $R$-CPE [($\tau = (RP)^{-N}$] are estimated from the equivalent circuit model for both light and dark impedance data \cite{kim2012persistent}. In ordered regions ( values are 0.19 millisecond and 95.35 nanosecond at 0 V and 3 V bias voltage; and in disordered regions the ( values are 0.25 and 0.23 microsecond at 0 V and 3 V bias voltage. When photo generated carriers are present, the ( values in ordered regions reduces to microsecond, and it further reduces under 3 V bias voltage; and in disordered regions the $\tau$ values do not vary significantly.
\begin{figure}[htp]
\centering
\includegraphics[scale=0.5]{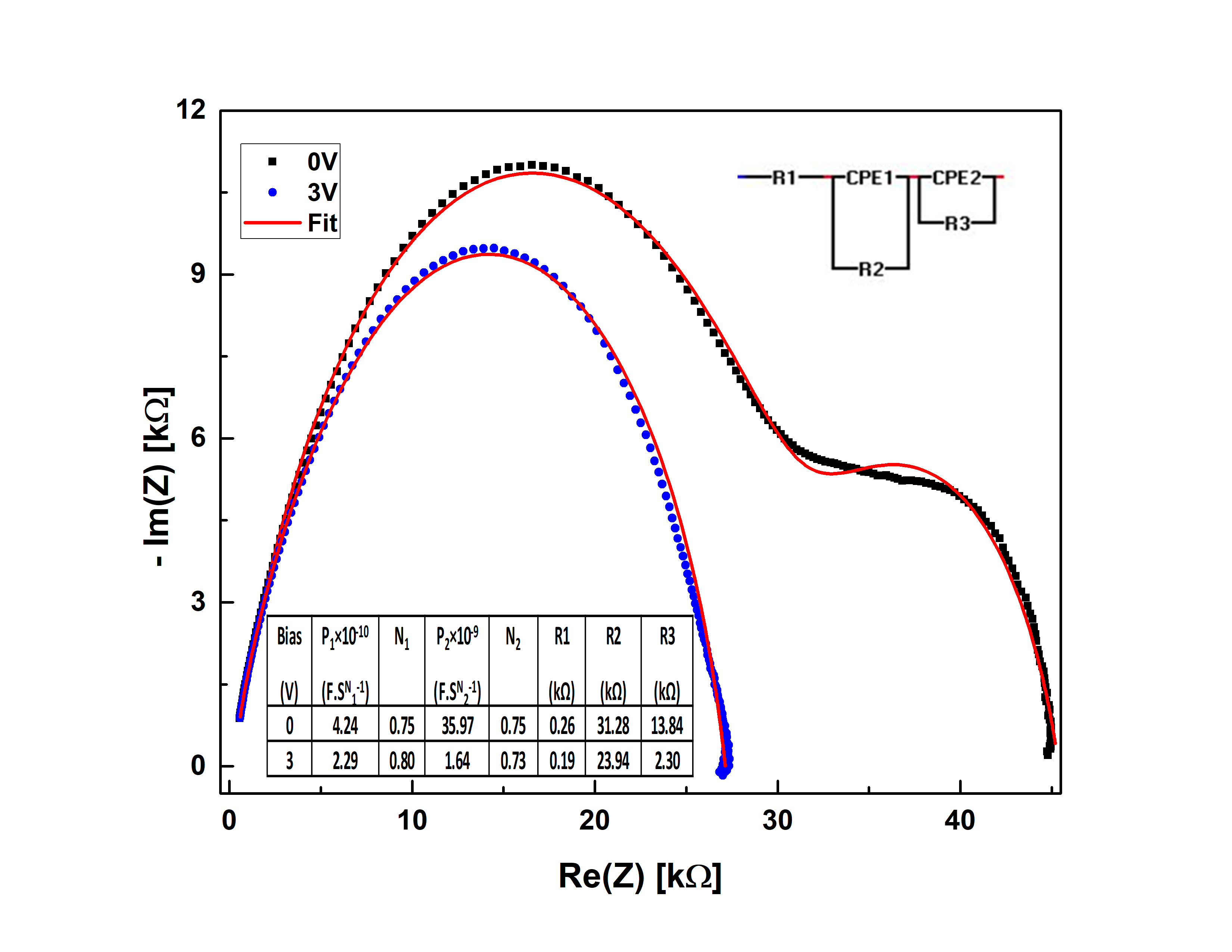}
\caption{Nyquist plots for P3HT at different voltages for doping-1 and fitted with equivalent circuit as shown in inset and the table shows the fit values to equation (\ref{equ 4}).}\label{fig. 6}
\end{figure}

\begin{figure}[htp]
\centering
\includegraphics[scale=1]{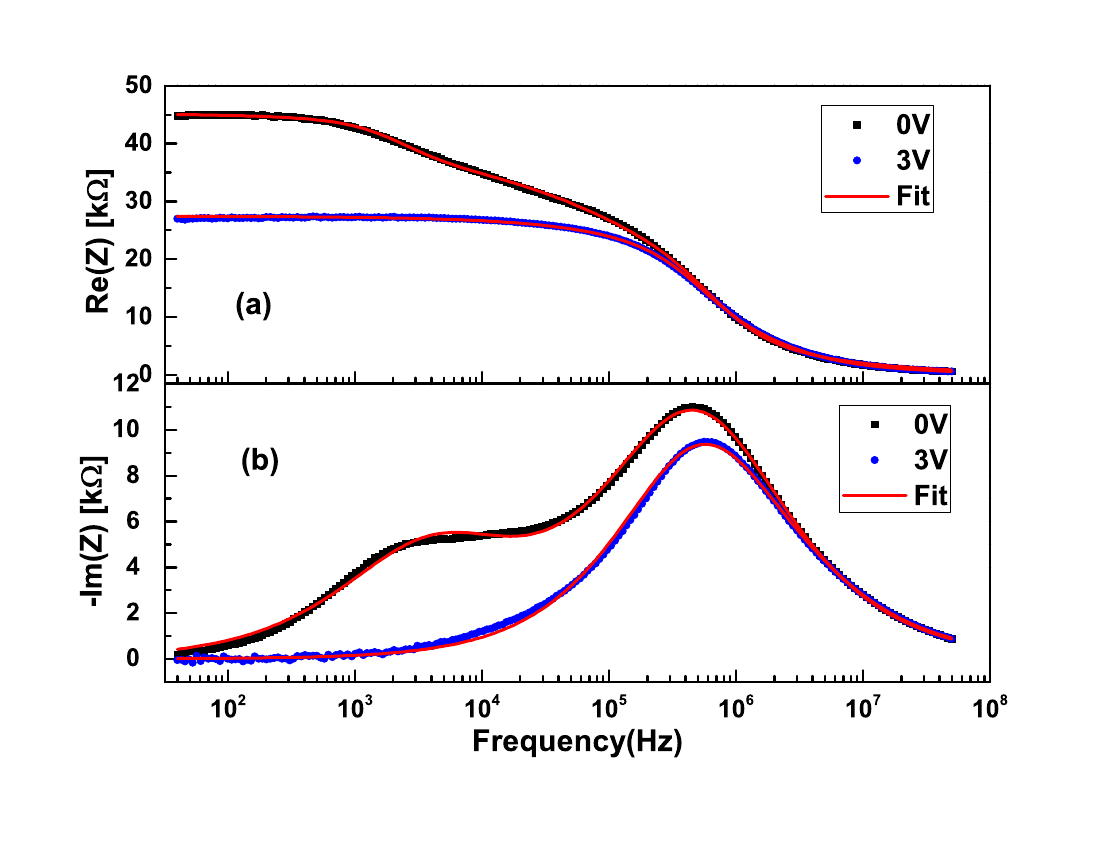}
\caption{(a) Real part of impedance vs frequency (b) Imaginary part of impedance vs frequency for doping-1. Data fits to equations (\ref{equ 5}) and (\ref{equ 6}).}\label{fig. 7}
\end{figure}

\begin{figure}[htp]
\centering
\includegraphics[scale=0.5]{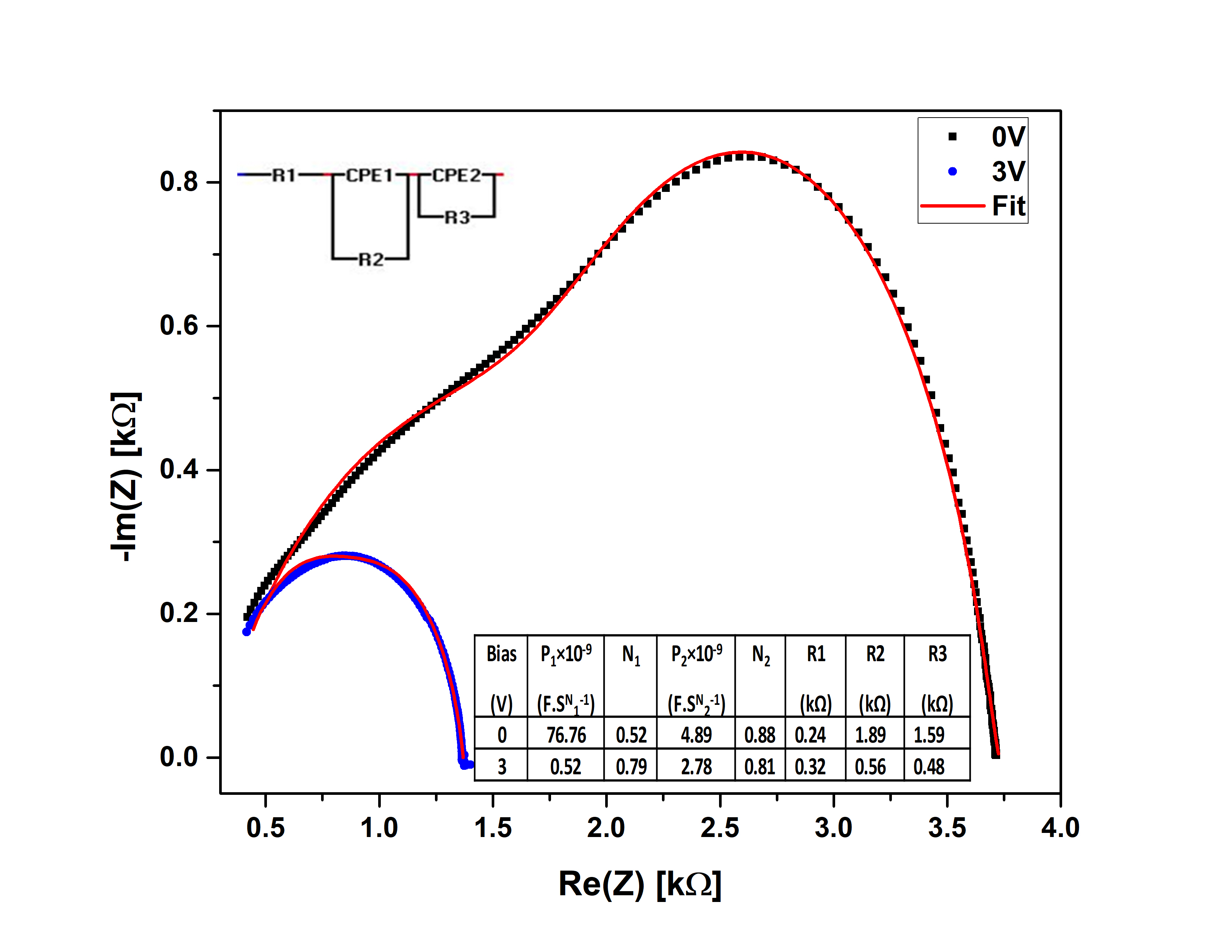}
\caption{Nyquist plots for P3HT at different voltages for doping-2 and fitted with equivalent circuit as shown in inset and the table shows the fit values to equation (\ref{equ 4}).}\label{fig. 8}
\end{figure}

It is well known that in semiconducting polymers, doping by the removal of an electron from chains by an oxidant, increases the electrical conductivity by serval orders of magnitude, also the carrier density and mobility increases \cite{singh2006structure,doi:10.1063/1.436503}. For example, even very low level of doping, the temperature dependence of conductivity varies significantly in these samples. The impedance data for chemically doped P3HT at the two different levels of doping i.e., doping-1 and doping-2 are shown in figures \ref{fig. 6} and \ref{fig. 8}. At doping-1, the impedance response shows one semicircle and one shoulder at zero bias. The complex impedance data is fitted by the equations as given by,	
\begin{eqnarray}
Z&=&R1+\frac{R2}{1+{(j\omega)}^{N_1}P_1R2}+\frac{R3}{1+{(j\omega)}^{N_2}P_2R3}\label{equ 4}
\\
Re(Z)&=&R1+\frac{R2\left(1+R2P_1\omega^{N_1}\cos{\frac{N_1\pi}{2}}\right)}{\left(1+R2P_1\omega^{N_1}\cos{\frac{N_1\pi}{2}\ }\right)^2+\left(R2P_1\omega^{N1}\sin{\frac{N_1\pi}{2}}\right)^2}\nonumber\\
&+&\frac{R3\left(1+R3P_2\omega^{N_2}\cos{\frac{N_2\pi}{2}}\right)}{\left(1+R3P_2\omega^{N_2}\cos{\frac{N_2\pi}{2}\ }\right)^2+\left(R3P_2\omega^{N_2}\sin{\frac{N_2\pi}{2}}\right)^2}\label{equ 5}
\\
-Im(Z)&=&\frac{{R2}^2P_1\omega^{N_1}\sin{\frac{N_1\pi}{2}}}{\left(1+R2P_1\omega^{N_1}\cos{\frac{N_1\pi}{2}\ }\right)^2+\left(R2P_1\omega^{N_1}\sin{\frac{N_1\pi}{2}}\right)^2}\nonumber\\
&+&\frac{{R3}^2P_2\omega^{N_2}\sin{\frac{N_2\pi}{2}}}{\left(1+R3P_2\omega^{N_2}\cos{\frac{N_2\pi}{2}\ }\right)^2+\left(R3P_2\omega^{N2}\sin{\frac{N_2\pi}{2}}\right)^2}\label{equ 6}
\end{eqnarray}

At doping-1, the values for fit parameters by using equation (\ref{equ 4}) is shown in inset of figure \ref{fig. 6}. The shoulder feature disappears at 3 V bias voltage. Frequency dependence of real part and imaginary parts are fitted by using equations (\ref{equ 5}) and (\ref{equ 6}), as shown in figure \ref{fig. 7}. These data show two relaxation processes at 0 V bias and one relaxation process at 3 V bias. As more carriers are injected at 3 V, the relaxation processes tend to become more averaged out throughout the sample. The obtained circuit element values are shown in the inset of figure (\ref{fig. 6}). The resistance values in the doped-1 sample are lower than in the case of undoped sample.

\begin{figure}[htp]
\centering
\includegraphics[scale=1]{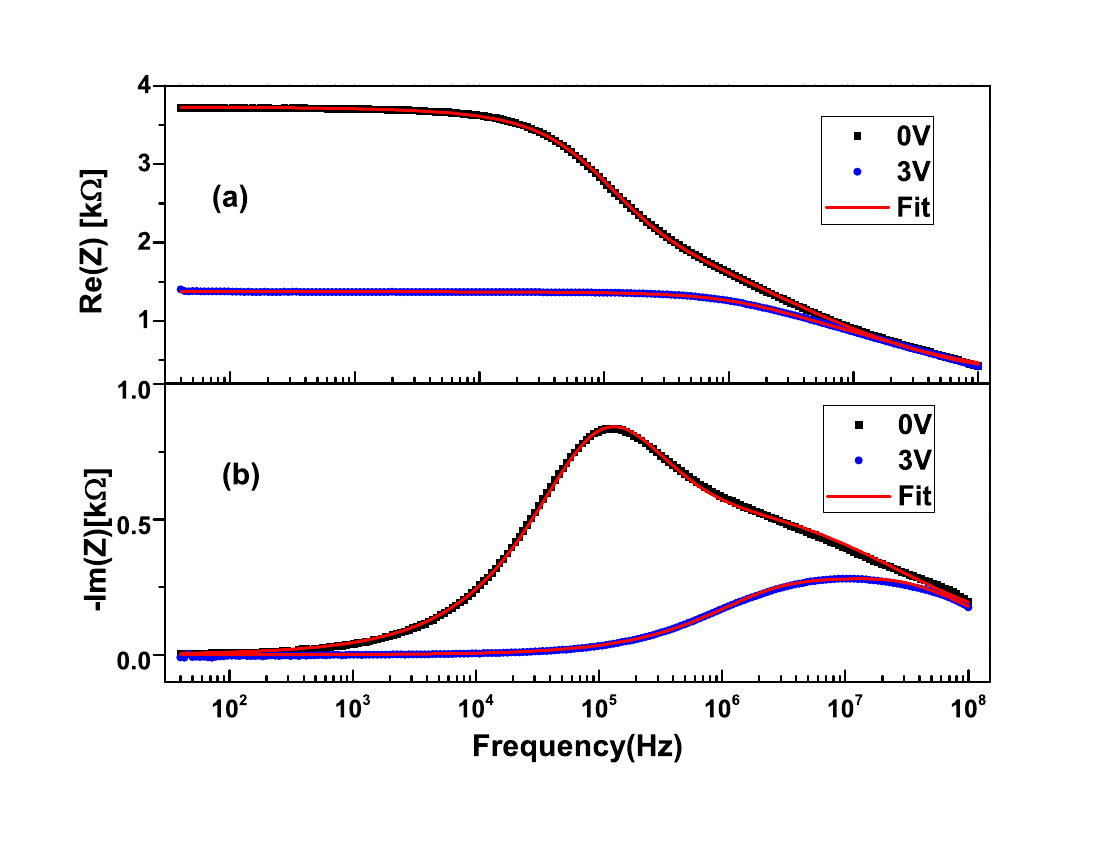}
\caption{Nyquist plots for P3HT at different voltages for doping-2 and fitted with equivalent circuit as shown in inset and the table shows the fit values to equations (\ref{equ 5}) and (\ref{equ 6}).}\label{fig. 9}
\end{figure}

Upon further increasing the doping level in same sample, the plot for impedance data, as in figure \ref{fig. 8}, show one broad semicircle;and at 3V bias it shrinks considerably.The data for the circuit elements, as in the inset of figure \ref{fig. 8}. In this case too, the data fit well to equations (\ref{equ 4})-(\ref{equ 6}). The resistance values are rather low, and its variation among each other is also less. The frequency dependence of real and imaginary parts, as in figure \ref{fig. 9}, show that as the carrier density further increases the variations in the relaxation process in different regions of the sample tend to narrow down to some typical value, and it is especially so at 3 V bias voltage.

\begin{figure}[htp]
\centering
\includegraphics[scale=0.4]{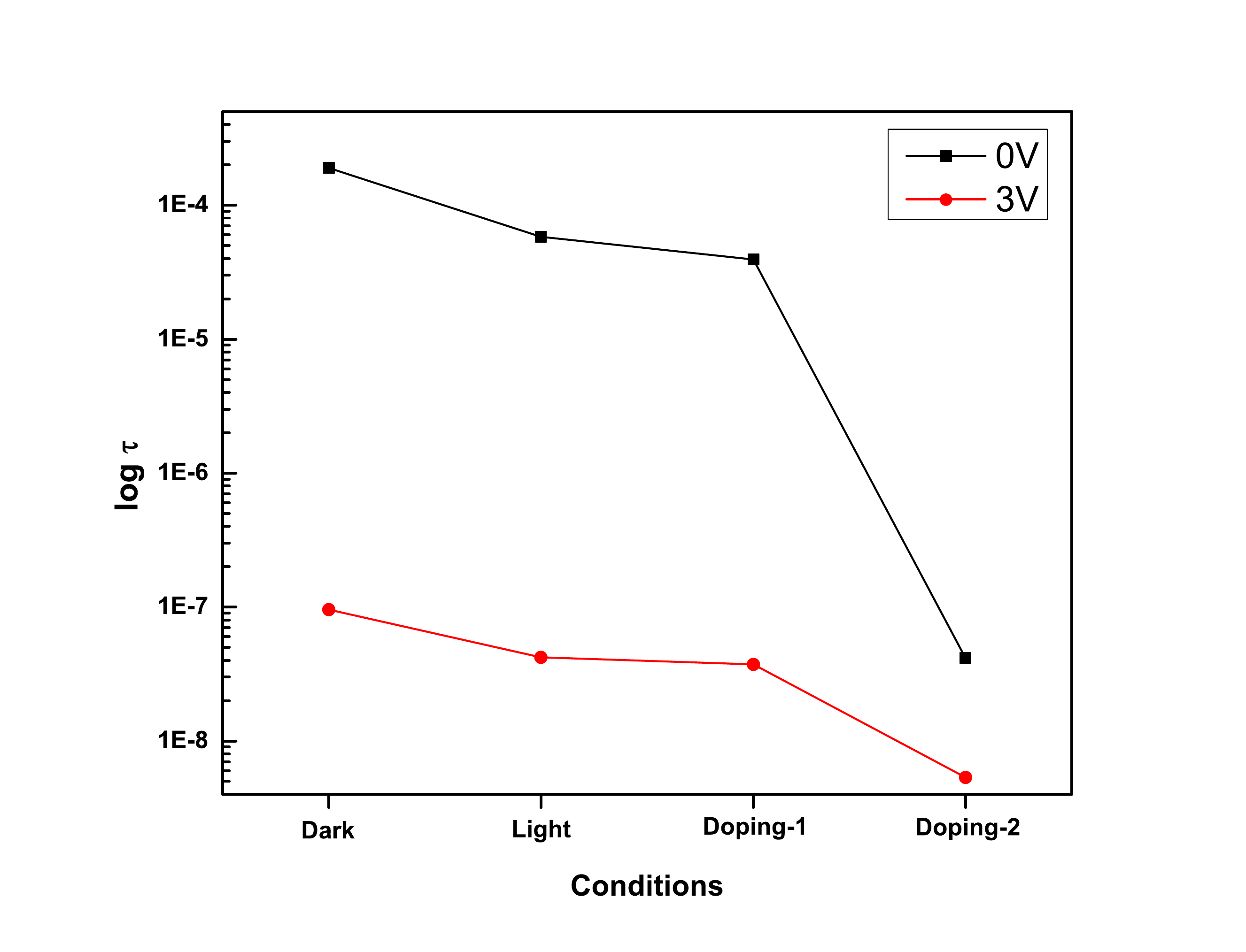}
\caption{The calculated relaxation time from the circuit elements in the ordered regions as a function of the variation in carrier density and bias voltage.}\label{fig. 10}
\end{figure}

Hence the controlled variations of charge density by applying bias voltage, light-induced charge carrier generations and chemical doping have significant effects on the impedance response. The impedance data for 1 V is included in the supplementary part, and the data for 2 and 3 V are not very different. The values of the circuit elements obtained from the fitting of data is used to estimate the relaxation times, and it varies by several orders of magnitude in the ordered regions. The plot of the relaxations time as a function of the variation in the carrier density is shown in figure \ref{fig. 10} at both 0 and 3V bias voltages; and under biased conditions its variation is less. Calculated values for the relaxation times in both ordered and disordered regions are in the table in the supplimentary file.

\section{Conclusions}

Impedance measurements in free-standing regioregular P3HT films, undoped, photo-doped and chemically doped, are carried out in this work. The data fit from Nyquist plots and the model consisting of resistive, capacitive and constant phase elements are used to find how the carrier transport and relaxation processes varies as a function of carrier density. As the carrier density increase the values of the resistive elements decreases considerably. Also, increase in carrier density tend to result in one dominant relaxation process, as observed from Nyquist plots.

\section{Acknowledgements}

SM would like to thank Amit, Arya, Dilip for help in lab and Subhashis for helpful discussions. SM acknowledge DST for INSPIRE research fellowship. Authors acknowledge department of physics, IISc for XRD measurement.

\section*{References}
\bibliographystyle{iop}
\bibliography{reference}

\end{document}